# Graphene re-knits its holes

Recep Zan[1,2], Quentin M. Ramasse[3,*], Ursel Bangert[2] and Konstantin S. Novoselov[1]

[1]School of Physics and Astronomy, The University of Manchester, Manchester, M13 9PL, United Kingdom

[2]School of Materials, The University of Manchester, Manchester, M13 9PL, United Kingdom

[3]SuperSTEM Laboratory, STFC Daresbury Campus, Daresbury WA4 4AD, United Kingdom

*To whom correspondence should be addressed: QMRamasse@superstem.org

**Abstract.** Nano-holes, etched under an electron beam at room temperature in single-layer graphene sheets as a result of their interaction with metal impurities, are shown to heal spontaneously by filling up with either non-hexagon, graphene-like, or perfect hexagon 2D structures. Scanning transmission electron microscopy was employed to capture the healing process and study atom-by-atom the re-grown structure. A combination of these nano-scale etching and re-knitting processes could lead to new graphene tailoring approaches.

**TOC Graphic.**

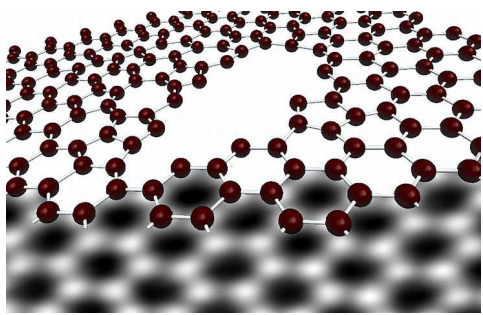

The development of graphene devices has entered a new era: graphene is rapidly moving from the laboratory to the factory floor with an increasing effort directed towards achieving devices that employ suspended membranes as well as graphene of defined geometries, e.g., nano-ribbons and quantum dots, because of their predicted novel, exceptional electronic properties.[1,2] As a result, the interfaces between graphene and other components of the devices such as metal contacts are under intense scrutiny. We have recently shown that in the presence of metals graphene can be etched on the nano-scale[3,4] under exposure to an electron probe following a sequence of point defect reactions, a process different from previous attempts at graphene etching. In earlier studies, graphene and graphite were etched or tailored in a gas environment (hydrogen or oxygen) at high temperatures.[5-8] In particular, the interaction of metals with the graphene surface in such extreme conditions was shown both experimentally[8] and theoretically[9] to provide an efficient means of controlling the etching process and effectively sculpt graphene/graphite nanostructures. A recent review of these techniques was for instance compiled by Biró and Lambin.[8] Additionally, a number of papers have previously discussed defects in graphene, either following experimental transmission electron microscopy (TEM) studies,[10-16] where defects were introduced by the electron beam, or by simulations based on Density Functional Theory (DFT) calculations.[17-24] Defects studies in all those papers are based on reconstruction of the graphene lattice as a result of knock-on of carbon atoms from the graphene lattice (mono- and multi- vacancy) or/and as a result of bond rotations, e.g., the formation of Stone-Wales (55-77) and 8-ring defects. Dynamics, stability and favourable arrangements of the reconstructed lattice, with the occurrence of larger vacancy aggregates and holes, as well as edge reconstructions thereof, are further aspects of those studies. Furthermore, new structures with long range order, such as haeckelites[18] consisting of 5-, 6- and 7-member rings, or pentaheptides[17] incorporating 5- and 7-member rings, have been proposed as result of theoretical calculations. Here we report the first observations of reconstruction, *i.e.* the mending and filling of many-vacancy holes (over 100 vacancies) in graphene, in an electron microscope. We show that provided a reservoir of loose carbon atoms is readily available nearby, holes in graphene can be re-filled with either non-hexagonal near-amorphous or perfectly hexagonal 2-dimensional structures.

To study atomic arrangements we employ high angle annular dark field (HAADF) imaging in a Nion UltraSTEM100™ aberration-corrected scanning transmission electron microscope (STEM). The microscope was operated in 'Gentle STEM' conditions[25] at 60 keV primary beam energy to prevent knock-on damage to the graphene sheets, even after long intervals of repeated scanning of small areas of a few square nanometers. The ultra-high-vacuum (UHV) design of this instrument allows clean imaging conditions with pressures below 5x10$^{-9}$ Torr near the sample. The beam was set up to a convergence semi-angle of 30 mrad with an estimated beam current of 45 pA. In these operating conditions the estimated probe size is 1.1 Å, providing the perfect tool for atom-by-atom chemical analysis[26]. These experimental conditions (scanning probe, low primary beam energy, high vacuum conditions) are significantly different from other studies of the dynamics of defects and edges in graphene, which employ higher beam energies and are typically carried out in much poorer vacuum conditions.[12,27,28] Samples were obtained by transfer of graphene membranes grown by CVD on copper substrates[29] to TEM grids. Two different metals were deposited by electron beam (Pd) and thermal (Ni) evaporation. Although there is no damage inferred to pristine graphene with a 60 keV electron beam, we have recently demonstrated[3,4] that in the presence of transition-metal or silicon atoms, hole formation does occurs under the e-beam. It is believed that the process occurs due to metal-catalysed dissociation of C-C bonds, which leads to point defect formation. The role of the electron beam in this is mainly to mobilise the metal atoms on graphene, and not to initiate the hole formation in the first place, i.e., by carbon atom displacement. The beam may also act as a local heat source and thus help overcome the defect formation energy barrier.[4] This process is however different from high-temperature TEM observations of graphene using a dedicated hot-stage holder, during which amorphous contaminations layers were shown to either be removed totally,[30] thus cleaning the graphene, or be transformed into crystalline patches.[31]

Here, the resulting defects are enlarged in the presence of further metal atoms, which have migrated under the electron beam to the edge of the incurred hole, upon which the process repeats itself. The nature of the intentionally-deposited metal impurities was confirmed by single-atom sensitive electron energy loss spectroscopy (EELS).[4] When the reservoir of metal atoms is exhausted, the process stops and the hole remains of almost the same size, although the edges undergo repeated reconstruction in consecutive scans. If metal impurities are not present and, at the same time, carbon atom supply is warranted, e.g. through near-by hydrocarbon contamination attracted towards the hole by the scanning probe, 'filling' occurs and the hole mends itself by non-hexagon arrangements. However, if no hydrocarbon contamination is present, healing can occur *via* reconstruction of the perfect graphene hexagon structure. The following Z-contrast images depict the hole evolution and two filling scenarios, by non-hexagon and by hexagon structures, in freely suspended graphene films where heavier foreign atoms are present either as residual impurities or as a consequence of metal evaporation.

The deposition of metal atoms on CVD-grown graphene results in the formation of nm-sized metal clusters sitting preferentially on patches of C-based surface contamination, ubiquitous in suspended graphene membranes.[32] Pd atoms, dislodged from a nearby Pd cluster by the electron beam and 'dragged' across the sample until they settle in a new stable position as the beam is being scanned, can be seen in fig. 1a to decorate a hole in graphene. Newly arriving Pd atoms lead to hole expansion (fig 1b), whereas in the absence of metal atoms the hole formation slows down and edge reconstruction takes place under the electron beam (fig. 1c). Furthermore, figure 1 shows that the hole is bordered by hydrocarbon contamination, which constitutes the carbon reservoir for the hole healing described below.

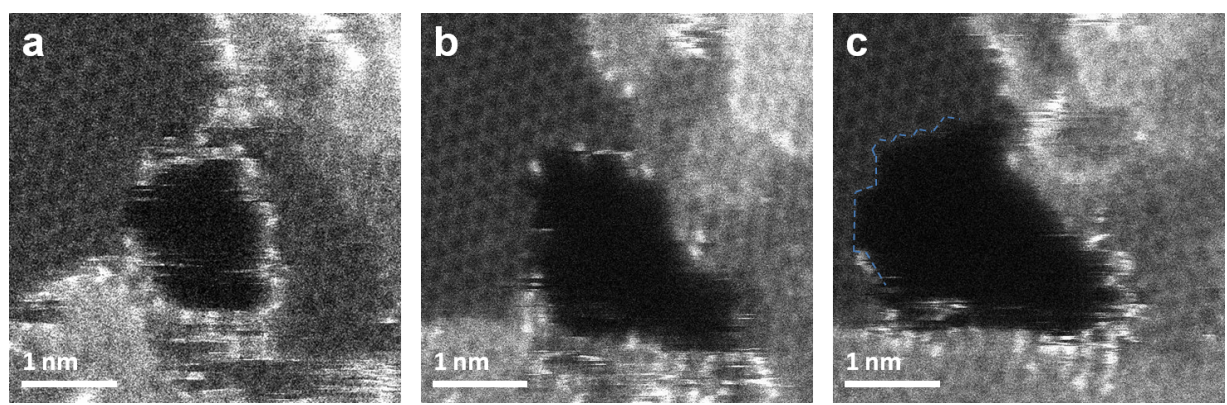

**Figure 1.** Atomic resolution HAADF images (raw data) from consecutive scans of suspended graphene, showing a) an etched hole decorated with Pd atoms; b) the enlargement of the hole following further supply of Pd atoms; c) the stabilization of the hole size in the absence of Pd atoms at the edge.

Figure 2 shows the same hole as in fig. 1. after the metal reservoir has been mostly exhausted. The hole then starts to heal with carbon atoms being supplied by the contamination patches (light grey features at the top and bottom of fig. 1 and 2). The image in figure 2a was taken after that in fig. 1c and illustrates the start of the healing process. The blue dotted line in fig. 2d shows the border of the hole at the pre-healing stage (immediately after fig. 1c was acquired). After a subsequent scan the hole has filled completely with what can be described as a 2-dimensional amorphous carbon patch (fig. 2b), constituted almost randomly of 5, 6-, 7- and even 8-member rings. Notably 5-7-/ 8-member ring pairs can arise from dissociation of lattice vacancies and constitute dislocation dipoles.[19] Additional impurity atoms are also captured within this newly-formed 'net': bright atoms are clearly seen on fig. 2a and 2b. Figure 2c is of a subsequent scan of the same area as in figs 2a and b. The deposition of carbon ad-atoms, which are dragged out of the hydrocarbon contamination by the scanning beam can be seen from slightly lighter patches and streaks on the carbon rings in fig. 2c, making atomic position identification slightly more difficult. Nevertheless, a redistribution of the defects is noticeable, indicating bond rotation during the scan: notably in fig. 2c the carbon atoms have re-organized themselves solely into 5-7 rings while the more unstable 8-member ring structure has disappeared. The bottom row images are the same as in the top row but in order to reduce the noise levels and to improve the accuracy of the image analysis (to show the carbon-atom 'network' more clearly), a probe deconvolution algorithm based on maximum entropy methods was applied to the raw Z-contrast images. This algorithm also has the advantage of removing the contribution of the tails of the focussed electron probe to the neighbouring atom intensities, which is essential for quantitative contrast analysis.[26,33] This sequence of images shows clearly that the faults in the reconstructed graphene do not arise from damage by the e-beam, but are grown-in features as a result of an 'epitaxial'-like growth. The incorporation of sets of 5-, 6- and 7-member rings structures, called Haeckelites, into highly defected graphene was suggested by Terrones et al.[18] but had never been observed experimentally hitherto. Figure 2 shows clearly that this kind of structure can exist in suspended form although they form here an amorphous patch rather than organising themselves into an ordered array of polygons.

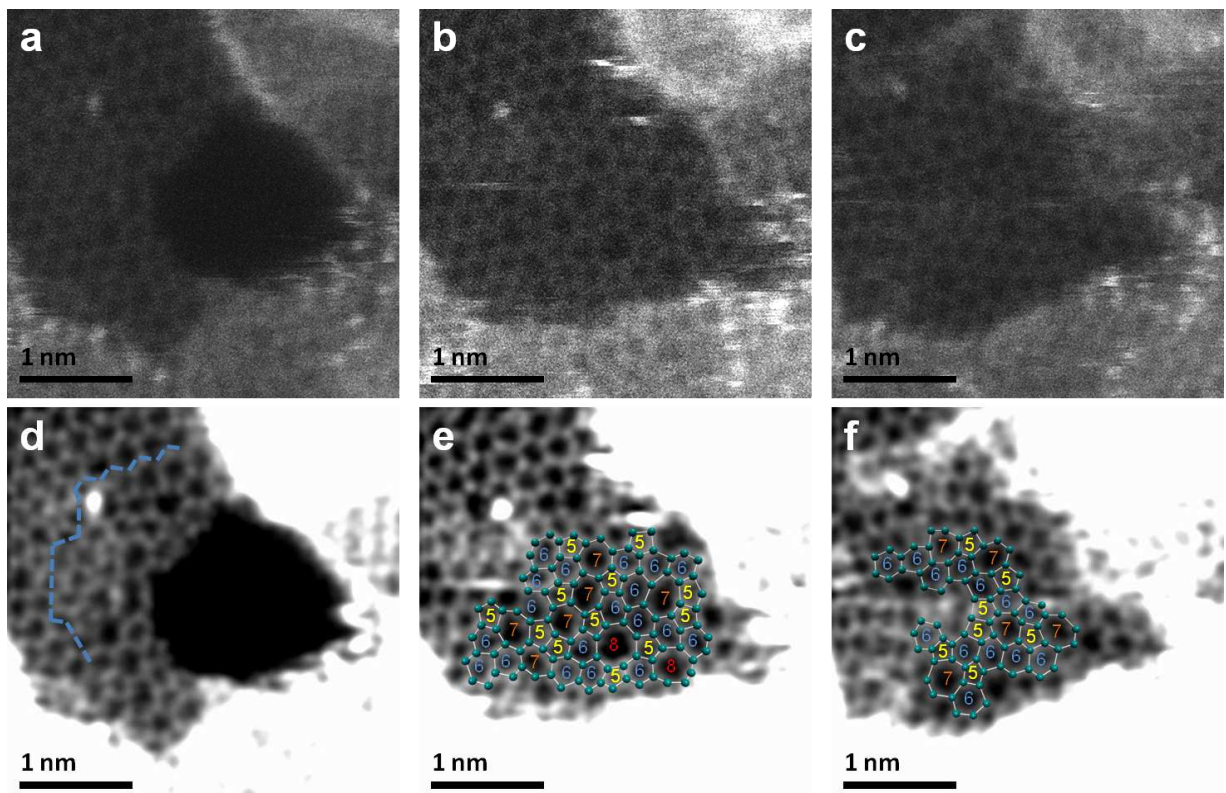

**Figure 2.** Atomic resolution Z-contrast images illustrating the hole filling process in suspended graphene. a) a hole created at the border of hydrocarbon contamination b) complete reconstruction with incorporation of 5-7 rings and two 5-8 rings, and c) redistribution of defects in the 'mended' region, by 5-7 rings. Images d-f are processed versions of a-c. A maximum entropy deconvolution algorithm was used and the contrast was optimised to visualise the carbon atoms. The carbon atom positions are highlighted by light green dots and polygons numbered according to the number of atoms in the rings.

Such 2-dimensional near-amorphous carbon layers can also grow 'epitaxially' on graphene, as illustrated on figure 3. After long observations of the same area of single layer graphene (more than 40 minutes of observation in this case) some contamination can occasionally 'creep in' and cover the graphene sheet. The circled region on figure 3a is a single additional carbon layer (as can be clearly deduced from the contrast in the image), which has formed on top of the clean, underlying (dark) graphene patch after repeated scanning. The presence of a Moiré pattern indicates that this additional layer has a similar structure to the underlying graphene, but the patter is highly irregular and interrupted (as highlighted by the arrows), unlike the patterns observed in pure Bernal-stacked graphene.[34]

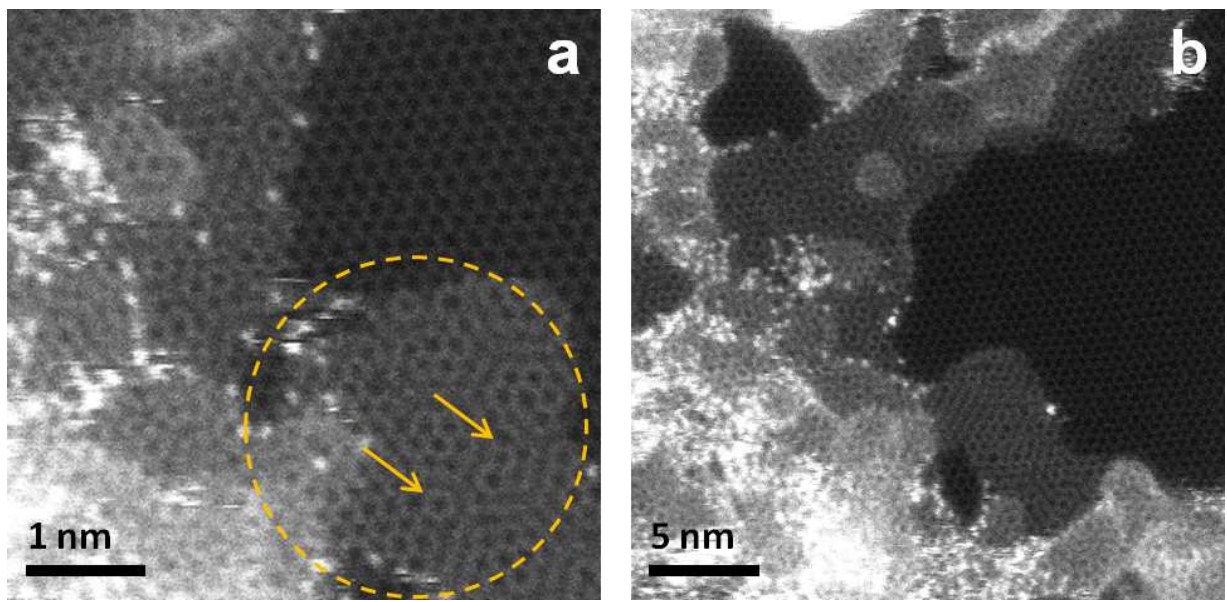

**Figure 3**. Atomic resolution HAADF image (raw data) obtained after repeated scanning of the same area; the circled area is an additional one-atom-thick carbon layer that has grown 'epitaxially' as a result of carbon atom supply from the hydrocarbon contamination surrounding the pristine graphene patch. The Moiré pattern, however, is non-regular (arrow), indicating the existence of random defected carbon ring networks in the newly-grown layer as well as in the already existing contamination.

The newly-grown layer must therefore be heavily faulted. It likely contains a high density of 5- and 7-member rings, similar to the 're-knitted' layer in fig 2b. This can again be interpreted as an amorphous 2-dimensional layer with some degree of short-range graphene-like order. The re-grown structures in figs. 2 and 3 provide a remarkable glimpse into the atomic arrangements of amorphous carbon structures; they show the precise atomic arrangements of a 2-D amorphous carbon material.[14] By close inspection of atomic resolution Z-contrast images of graphene films it can then be concluded that a large fraction of the ubiquitous contamination (such as the lighter patches in fig. 3b) where material deposited on graphene sits preferentially, consists of stacks of these amorphous mono-layer carbon films.

By contrast, when a small hole is created away from the hydrocarbon contamination, the self-healing mechanism appears to lead to the formation of perfect hexagons. Two holes are apparent in fig. 4a, both created as a result of Ni-atom mediated graphene etching. The small hole was created when Ni atoms from the edge of the larger hole, which had formed earlier, were dislodged by the electron beam and 'dragged' onto a clean patch of graphene where the etching was initiated, through the e-beam scanning motion. In subsequent scans, after cessation of the migration of metal atoms to its edges, the small hole is observed to fill up with perfect carbon hexagons (fig. 4b). This 're-knitting' process happened within a much shorter time span (less than 10 seconds: two consecutive scans) compared to the previous case (figs. 1&2). Impurity atoms are again trapped within the newly formed lattice, such as the brighter atom indicated in fig. 4b by an arrow. After the contribution of the electron probe 'tails' to the raw HAADF images has been removed (using for instance a maximum-entropy-based probe deconvolution algorithm[33]), a careful comparison of atomic intensities vs. atomic number Z with the model value of $Z^{1.7}$ for HAADF imaging (in our experimental conditions) can enable the chemical identification of these impurities.[26] Figure 4c shows a histogram of the frequency of occurrence of atoms versus their HAADF intensity in the vicinity of the bright atom in Fig. 4b, obtained from the deconvoluted image in the inset in fig. 4c. The positions of the intensity maxima are in the ratio of 1:1.66, which is close to the $Z^{1.7}$ ratios for Z = 6 (C) and Z = 8 (O), of 1:1.63. Although there is only one such impurity in this image (and the statistics are therefore low) this is a good indication that the brighter atom is likely oxygen.

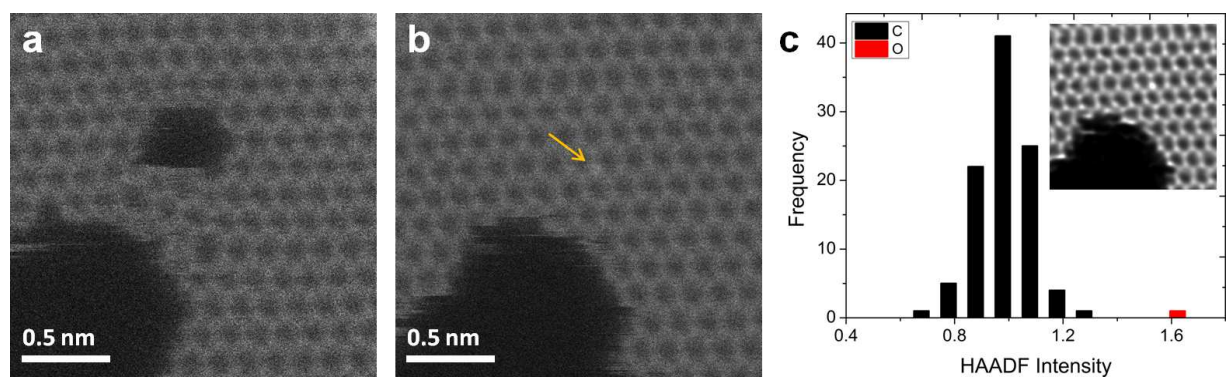

**Figure 4.** Atomic resolution HAADF images (raw data) from subsequent scans of suspended graphene, showing a) two holes as result of Ni-mediated etching, b) complete reconstruction of small hole with perfect hexagon graphene structure. c) histogram of the atom intensities, in the vicinity of the brighter atom: the analysis was carried out after processing image 4b with a probe deconvolution algorithm (inset).

The size of the hole to fill is an obvious difference with the previous case and may explain the formation of perfect hexagons: arguably, defects are less likely to form across such a short growth distance. Moreover, the larger distance from the contamination reservoir and the proximity to a much larger perforation in the graphene sheet may also play a role. Edges in graphene are known to be more unstable even at low voltages[12,15]: atoms ejected from the edge of this larger neighbouring hole may have been captured during the re-knitting while the long edge may have reconfigured to accommodate the newly formed material into the usual, energetically favourable honeycomb pattern. Additionally, using a low primary beam energy (60 keV), clean vacuum conditions (<$5 \times 10^{-9}$ Torr) to reduce the ionization damage probability, and a scanning probe, can help explaining our observations of hole-filling, which contrast with previous reports of hole-expansion at 80 keV with a stationary beam.[12,28]

In conclusion, etching of graphene occurs in the presence of metals; when the metal atom supply ceases graphene reconstructs its holes by forming perfect hexagonal or polygonal, *i.e.* 5-, 6-, 7- and 8-member ring structures, which can exist either in suspended form or grow 'epitaxially' on top of clean graphene. Carbon atoms knocked out from neighbouring edges, or supplied by nearby hydrocarbon contamination patches, are incorporated into the holes in a remarkable re-knitting process. This observation of self-healing of graphene might open up possibilities for the use of e-beam techniques in tailoring graphene in a 'bottom-up' fashion *via* nano-scale-controlled etching and epitaxial growth from sheet edges.

**References**

1.	Geim, A. K.; Novoselov, K. S. *Nat Mater* **2007,** 6, (3), 183-191.

2.	Ponomarenko, L. A.; Geim, A. K.; Zhukov, A. A.; Jalil, R.; Morozov, S. V.; Novoselov, K. S.; Grigorieva, I. V.; Hill, E. H.; Cheianov, V. V.; Falko, V. I.; Watanabe, K.; Taniguchi, T.; Gorbachev, R. V. *Nat Phys* **2011,** 7, (12), 958-961.

3.	Zan, R.; Bangert, U.; Ramasse, Q.; Novoselov, K. S. *Journal of Physical Chemistry Letters* **2012,** 3, 953-958.

4.	Ramasse, Q. M.; Zan, R.; Bangert, U.; Boukhvalov, D. W.; Son, Y. W.; Novoselov, K. S. *ACS Nano* **2012**, DOI: 10.1021/nn300452y.

5.	Severin, N.; Kirstein, S.; Sokolov, I. M.; Rabe, J. P. *Nano Letters* **2008,** 9, (1), 457-461.

6.	Booth, T. J.; Pizzoccero, F.; Andersen, H.; Hansen, T. W.; Wagner, J. B.; Jinschek, J. R.; Dunin-Borkowski, R. E.; Hansen, O.; Boggild, P. *Nano Letters* **2011**, 11, (7), 2689-2692.

7.	Dimiev, A.; Kosynkin, D. V.; Sinitskii, A.; Slesarev, A.; Sun, Z.; Tour, J. M. *Science* **2011,** 331, (6021), 1168-1172.

8.      Biró, L. P.; Lambin, P. *Carbon* **2010,** 48, (10), 2677-2689.

9.      Chan, K. T.; Neaton, J. B.; Cohen, M. L. *Physical Review B* **2008,** 77, (23), 235430.

10.     Hashimoto, A.; Suenaga, K.; Gloter, A.; Urita, K.; Iijima, S. *Nature* **2004,** 430, (7002), 870-873.

11.     Meyer, J. C.; Kisielowski, C.; Erni, R.; Rossell, M. D.; Crommie, M. F.; Zettl, A. *Nano Letters* **2008,** 8, (11), 3582-3586.

12.     Girit, C. O.; Meyer, J. C.; Erni, R.; Rossell, M. D.; Kisielowski, C.; Yang, L.; Park, C.-H.; Crommie, M. F.; Cohen, M. L.; Louie, S. G.; Zettl, A. *Science* **2009,** 323, (5922), 1705-1708.

13.     Banhart, F.; Kotakoski, J.; Krasheninnikov, A. V. *ACS Nano* **2011,** 5, (1), 26-41.

14.     Kotakoski, J.; Krasheninnikov, A. V.; Kaiser, U.; Meyer, J. C. *Physical Review Letters* **2011,** 106, (10), 105505.

15.     Kotakoski, J.; Meyer, J. C.; Kurasch, S.; Santos-Cottin, D.; Kaiser, U.; Krasheninnikov, A. V. *Physical Review B* **2011,** 83, (24), 245420.

16.     Gass, M. H.; Bangert, U.; Bleloch, A. L.; Wang, P.; Nair, R. R.; Geim, A. K. *Nat Nano* **2008,** 3, (11), 676-681.

17.     Crespi, V. H.; Benedict, L. X.; Cohen, M. L.; Louie, S. G. *Physical Review B* **1996,** 53, (20), R13303-R13305.

18.     Terrones, H.; Terrones, M.; Hernandez, E.; Grobert, N.; Charlier, J. C.; Ajayan, P. M. *Physical Review Letters* **2000,** 84, (8), 1716-1719.

19.     Ewels, C. P.; Heggie, M. I.; Briddon, P. R. *Chemical Physics Letters* **2002,** 351, (3-4), 178-182.

20.     Lee, G.-D.; Wang, C. Z.; Yoon, E.; Hwang, N.-M.; Kim, D.-Y.; Ho, K. M. *Physical Review Letters* **2005,** 95, (20), 205501.

21.     Krasheninnikov, A. V.; Banhart, F. *Nat Mater* **2007,** 6, (10), 723-733.

22.     Jeong, B. W.; Ihm, J.; Lee, G.-D. *Physical Review B* **2008,** 78, (16), 165403.

23.     Lusk, M. T.; Carr, L. D. *Physical Review Letters* **2008,** 100, (17), 175503.

24.     Kim, Y.; Ihm, J.; Yoon, E.; Lee, G.-D. *Physical Review B* **2011,** 84, (7), 075445.

25.     Krivanek, O. L.; Dellby, N.; Murfitt, M. F.; Chisholm, M. F.; Pennycook, T. J.; Suenaga, K.; Nicolosi, V. *Ultramicroscopy* **2010,** 110, (8), 935-945.

26.     Krivanek, O. L.; Chisholm, M. F.; Nicolosi, V.; Pennycook, T. J.; Corbin, G. J.; Dellby, N.; Murfitt, M. F.; Own, C. S.; Szilagyi, Z. S.; Oxley, M. P.; Pantelides, S. T.; Pennycook, S. J. *Nature* **2010,** 464, (7288), 571-574.

27.     Warner, J. H.; Rummeli, M. H.; Ge, L.; Gemming, T.; Montanari, B.; Harrison, N. M.; Buchner, B.; Briggs, G. A. D. *Nat Nano* **2009,** 4, (8), 500-504.

28.     Russo, C. J.; Golovchenko, J. A. *Proceedings of the National Academy of Sciences* **2012**, DOI: 10.1073/pnas.1119827109.

29.     Li, X.; Cai, W.; An, J.; Kim, S.; Nah, J.; Yang, D.; Piner, R.; Velamakanni, A.; Jung, I.; Tutuc, E.; Banerjee, S. K.; Colombo, L.; Ruoff, R. S. *Science* **2009,** 324, (5932), 1312-1314.

30.     van Dorp, W. F.; Zhang, X.; Feringa, B. L.; Wagner, J. B.; Hansen, T. W.; De Hosson, J. Th. M. *Nanotechnology* **2011,** 22, 505303.

31.     Barreiro, A.; Boerrnert, F.; Avdoshenko, S. M.; Rellinghaus, B.; Cuniberti, G.; Ruemmeli, M. H.; Vandersypen, L. M. K. Unpublished **2012**.

32.     Zan, R.; Bangert, U.; Ramasse, Q.; Novoselov, K. S. *Nano Letters* **2011,** 11, (3), 1087-1092.

33.     Ishizuka, K.; Abe, E. *Proc. of the 13th European Microscopy Congress, Instrumentation and Methodology* **2004,** 1, 117.

34.     Zan, R.; Bangert, U.; Ramasse, Q.; Novoselov, K. S. *Journal of Microscopy* **2011,** 244, (2), 152-158.